\documentclass[10pt,twocolumn]{article}
\usepackage{latex8,latexsym,epsfig}
\usepackage{times}

\newcommand{\C}[0]{{\mathbf{C}}}

\newcommand{\mat}[4]{\left(\begin{array}{rr}#1 & #2\\ #3 & #4 
\end{array}\right)
}

\newcommand{\proof}{\noindent\textit{Proof.} } 
\newcommand{\diag}{\mathop{\rm diag}\nolimits}

\newcommand{\nix}[1]{}

\def\ket#1{\left|#1\right>}
\def\qed{\quad{$\Box$}}

\def\IrrP(#1){\mathop{{\rm Irr}}\nolimits(\phi\,|\,#1)}
\newcommand{\onemat}[0]{\mbox{\bf 1}}

\newcommand{\DCT}[0]{\mbox{\rm DCT}}
\newcommand{\DST}[0]{\mbox{\rm DST}}

\newcommand{\ds}{\displaystyle}
\newtheorem{theorem}{Theorem}

\newtheorem{lemma}[theorem]{Lemma}

\begin{document}
\title{Discrete Cosine Transforms on Quantum Computers}
\author{
Andreas Klappenecker\\
Texas A\&M University\\
Department of Computer Science\\ 
College Station, TX 77843-3112, USA
\and
Martin R\"otteler\\
Universit{\"a}t Karlsruhe\\
Institut f{\"u}r Algorithmen und Kognitive Systeme\\
Quantum Computing Group, Prof.\ Thomas Beth\\
Am Fasanengarten 5, 76\,128 Karlsruhe, Germany}

\maketitle
\begin{abstract}
A classical computer does not allow to calculate a discrete cosine
transform on $N$ points in less than linear time.  This trivial lower
bound is no longer valid for a computer that takes advantage of
quantum mechanical superposition, entanglement, and interference
principles. In fact, we show that it is possible to realize the
discrete cosine transforms and the discrete sine transforms of size
$N\times N$ and types I,II,III, and IV with as little as $O(\log^2\!
N)$ operations on a quantum computer, whereas the known fast
algorithms on a classical computer need $O(N\log N)$ operations.
\end{abstract}

\section{Introduction}
Feynman proposed in 1982 a computational model that was based on the
principles of quantum physics instead of classical physics.  The model
has been considered a mere curiosity until Peter Shor showed in 1994
that it is possible to factor integers in polynomial time on a quantum
computer~\cite{Shor:94}. Thus, a moderate sized quantum computer is
for instance able to break the RSA public key cryptosystem.  Quantum
computing is an exciting area of emerging signal processing
applications.  In fact, signal processing methods play a key role in
Shor's integer factoring algorithm and in many other quantum
algorithms. 

A quantum computer is based on the concept of a quantum bit, just as a
classical computer is based on the notion of a bit.  A single quantum
bit represents the state of a two-level quantum system such as a
polarized photon or a spin-1/2 system.  Unlike a classical computer,
adding another quantum bit to the memory of a quantum computer will
not increase the dimensionality of the state space by one but will
\textsl{double} it, allowing for linear combinations 
of $2^n$ different base states in the case of $n$ quantum bits. 

A program on a quantum computer is composed of a sequence of
elementary `gates', which perform simple unitary transforms (as
explained in Section 3).  In fact, many algorithms in quantum
computing rely on the fast Fourier transforms, the Walsh-Hadamard
transforms, or other unitary transforms well-known in signal
processing. 

The purpose of this paper is to derive (extremely) fast quantum
algorithms for the discrete cosine and sine transforms.  These
algorithms can be implemented on a number of quantum computing
technologies based on Raman-coupled low-energy states of trapped
ions~\cite{cirac95,nagerl98}, nuclear spins in silicon~\cite{kane98}, 
electron spins in quantum dots~\cite{imamoglu99}, atomic cavity quantum
electrodynamics~\cite{rauschenbeutel99}, or linear
optics~\cite{knill01}.  Coherent control of up to four quantum bits
has been demonstrated~\cite{nist01}, and much progress is expected in
the near future.

\section{Definitions}
Recall the definitions~\cite{RY:90} of 
the {\em discrete cosine transforms}:
\[
\setlength{\arraycolsep}{0.2em}%
\begin{array}{l@{:=}l}
C_N^{\rm I} & \left(\ds\frac{2}{N}\right)^{1/2} 
\left[k_i \cos{\ds\frac{i j\pi}{N}}\right]_
   {i, j = 0..N}
\\
C_N^{\rm II} & \left(\ds\frac{2}{N}\right)^{1/2} 
\left[k_i \cos{\ds\frac{i (j+1/2)\pi}{N}}\right]_
   {i, j = 0..N-1}
\\
C_N^{\rm III} & \left(\ds\frac{2}{N}\right)^{1/2} 
\left[k_i \cos{\ds\frac{(i+1/2) j\pi}{N}}\right]_
   {i, j = 0..N-1}
\\
C_N^{\rm IV} & \left(\ds\frac{2}{N}\right)^{1/2} 
\left[k_i \cos{\ds\frac{(i+1/2) (j+1/2)\pi}{N}}\right]_
   {i, j = 0..N-1}
\end{array}
\]
where $k_i:=1$ for $i=1,\dots,N-1$ and $k_0 := {1}/{\sqrt{2}}$.
The numbers $k_i$ ensure that the transforms are orthogonal.
The discrete sine transforms $S_N^{\rm I}$, $S_N^{\rm II}$,
$S_N^{\rm III}$, and $S_N^{\rm IV}$
are defined accordingly, see \cite{RY:90}
for details.  Notice that 
$C_N^{\rm III}$ (resp. $S_N^{\rm III}$) is the transpose of 
$C_N^{\rm II}$ (resp. $S_N^{\rm II})$, hence it suffices to derive
circuits for the type II transforms. In the following, we content
ourselves to $N=2^n$, which is justified by the machine model introduced
below. 

\section{Quantum Gates}
A quantum computer consists of a system of $n$ two-level quantum
systems each of which represents a quantum bit (qubits). The computational
state space is given by $\C^{2^n}$.  Denote an orthonormal basis of
$\C^{2^n}$ by $\ket{x}$ where $x$ is an $n$-bit integer. The state of
the quantum computer can be manipulated by quantum gates. Two types of
operations are considered as elementary: the single qubit operations
and the controlled NOT operations~\cite{BBCshort:95}.

The single qubit operations are given by local unitary operations of
the form $ \onemat_{2^{n-t}}\otimes U\otimes \onemat_{2^{t-1}},$ with
$U\in {\cal U}(2)$. The controlled NOT gate operates on two qubits. It
negates the target qubit if and only if the control qubit is 1.  All
operations on a quantum computer can be built up from these elementary
operations, i.e., they form a universal set of gates. Therefore a
basic (and nontrivial) task is to find efficient factorizations for
explicitly given transformations into elementary gates

There is a graphical notation for quantum gates that has been
introduced by Feynman. Each line denotes a qubit with the most
significant bit on top. The circuits are read from left to right like
musical scores. The figure on the left shows a single qubit operation
$U\otimes \onemat_2$. The figure on the right shows a controlled NOT
operation, that is, the unitary transform $\ket{00}\mapsto\ket{00}$,  
$\ket{01}\mapsto\ket{11}$, 
$\ket{10}\mapsto\ket{10}$,  $\ket{11}\mapsto\ket{01}$.

\centerline{\input{test.pic}}
We refer the reader to \cite{BBCshort:95} for more information
about quantum gates and the graphical notation.

\section{DCT and DST of Type I}
We derive the circuits for the discrete sine and cosine transforms of
type I all at once. Indeed, the $\DST_{\rm I}$ and $\DCT_{\rm I}$ can be recovered from
the $\mbox{DFT}$ by a base change~\cite{wickerhauser93}
\begin{equation}\label{dctI}
T_N^\dagger \cdot F_{2N} \cdot T_N = 
C_N^{\rm I} \oplus i S_N^{\rm I},
\end{equation}
where
\[
\renewcommand{\arraystretch}{1.0}%
\setlength{\arraycolsep}{0.3em}%
T_N =
\left(
\begin{array}{rrrrrrrr}
1 & & & & & & & \\
  & \frac{1}{\sqrt{2}} & & & & \frac{i}{\sqrt{2}} & &  \\
  & & \begin{picture}(5,5) 
\multiput(0,4)(2,-2){3}{\makebox(0,0){.}}
\end{picture} & & & & \begin{picture}(5,5) 
\multiput(0,4)(2,-2){3}{\makebox(0,0){.}}
\end{picture} & \\
  & & & \frac{1}{\sqrt{2}} & & & & \frac{i}{\sqrt{2}}\\
 & & & & \;1 & & & \\
  & & & \frac{1}{\sqrt{2}} & & & & {-} \frac{i}{\sqrt{2}}\\
  & &  \begin{picture}(5,5) 
\multiput(0,0)(2,2){3}{\makebox(0,0){.}}
\end{picture} & & & & \begin{picture}(5,5) 
\multiput(0,0)(2,2){3}{\makebox(0,0){.}}
\end{picture} &\\
  & \frac{1}{\sqrt{2}} & & & & {-} \frac{i}{\sqrt{2}} & &
\end{array}
\right)
\]
and $F_N =\frac{1}{\sqrt{N}} [ \exp(2\pi i\, kl/N) ]_{k,l=0..N-1}$
with $i^2=-1$ denotes the DFT of length $N$.  Since efficient quantum
circuits for the DFT are known~\cite{Shor:94}, it remains to find an
efficient implementation of the base change matrix $T_N$.

Denote the basis vectors of $\C^{2^{n+1}}$ by $\ket{b x}$, where $b$
is a single bit and $x$ is an $n$-bit number.  The two's complement of
an $n$-bit unsigned integer $x$ is denoted by $x'$, that is,
$x'=2^n-x$.  The action of $T_N$ can be described by
$$
\begin{array}{l@{\,}c@{\,}l@{\quad}l@{\,}c@{\,}l}
T_N \ket{0\mathbf{0}}&=&\ket{0\mathbf{0}}, &
T_N\ket{0x}&=&\frac{1}{\sqrt{2}}\ket{0x}+\frac{1}{\sqrt{2}}\ket{1x'},\\[1ex]
T_N \ket{1\mathbf{0}}&=&\ket{1\mathbf{0}}, &
T_N\ket{1x}&=&\frac{i}{\sqrt{2}}\ket{0x}-\frac{i}{\sqrt{2}}\ket{1x'},
\end{array}
$$
for all integers $x$ in the range $1\le x<2^n$, where $i^2=-1$. 
Ignoring the two's complement in $T_N$, we can define an operator $D$
by 
$$
\begin{array}{l@{\,}c@{\,}l@{\quad}l@{\,}c@{\,}l}
D\ket{0\mathbf{0}}&=&\ket{0\mathbf{0}}, &
D\ket{0x}&=&\frac{1}{\sqrt{2}}\ket{0x}+\frac{1}{\sqrt{2}}\ket{1x},\\[1ex]
D \ket{1\mathbf{0}}&=&\ket{1\mathbf{0}}, &
D\ket{1x}&=&\frac{i}{\sqrt{2}}\ket{0x}-\frac{i}{\sqrt{2}}\ket{1x},
\end{array}
$$
for all integers $x$ in the range $1\le x<2^n$. This operator is
essentially block diagonal and easy to implement by a single qubit
operation, followed by a correction. Indeed, define the matrix $B$ by $B =
\ds\frac{1}{\sqrt{2}}\mat{1}{i}{1}{-i}$, then Figure~\ref{dctIfact}
gives an implementation of the operator $D$. 
\begin{figure}[ht]
\vspace{2ex}
\input{dctIfact.pic}
\caption{\label{DCTIfact} Circuits for the
matrix $D$ and the permutation
$\pi$. 
}\label{dctIfact}
\end{figure}

Define $\pi$ to be the permutation given by a two's complement
conditioned on the most significant bit $\pi\ket{0x}=\ket{0x}$ and
$\pi\ket{1x}=\ket{1x'}$ for all $n$-bit integers $x$.  It is clear
that $T_N=\pi D$. The circuit for the permutation $\pi$ is shown in
Figure~\ref{dctIfact}. Here $P_n$ denotes the map 
$\ket{x}\mapsto \ket{x{+}1 \bmod 2^n}$ on $n$ qubits, see~\cite{PRBshort:99}
for an implementation. 

\begin{theorem}
The discrete cosine transform $C_N^{\rm I}$ and the discrete sine
transform $S_N^{\rm I}$ can be realized with $O(\log^2\!N)$ elementary
quantum gates; the quantum circuit for these transforms 
is shown in Figure~\ref{DCTIcomplete}. 
\end{theorem}
\begin{figure*}[ht]
\input{dct1.pic}
\caption{\label{DCTIcomplete}Complete quantum circuit for the $\DCT_{\rm
I}$}
\end{figure*}
\proof Let $N=2^n$. 
We note that $O(\log^2 N)$ quantum gates are sufficient to realize the DFT of
length $2N$, see~\cite{Shor:94}.  The permutation $\pi$ can be implemented with at most
$O(\log^2 N)$ elementary gates. At most 
$O(\log N)$ quantum gates are needed to realize the
operator $D$. This shows that the $\DCT_{\rm I}$ and the $\DST_{\rm I}$
can be realized with $O(\log^2 N)$ elementary quantum gates. The preceding
discussion shows that Figure~\ref{DCTIcomplete} realizes the 
$\DCT_{\rm I}$ and $\DST_{\rm I}.$~\qed

\section{DCT and DST of Type IV}

The trigonometric transforms of type IV are derived from the DFT by 
\begin{equation}\label{dctIV}
e^{\pi i /4N} R_N^t \cdot F_{2N} \cdot R_N = C_N^{\rm IV}
\oplus (-i)S_N^{\rm IV}.
\end{equation}
Here $R_N$ denotes the matrix 
$$
\begin{array}{l@{\,}c@{\,}l}
R_N & = & {\ds\frac{1}{\sqrt{2}}} \left(
\begin{array}{@{\,}c@{\,}c@{\,}c@{\,}c@{\,}c@{\,}c@{\,}c@{\,}c}
1 & & & & -i & & & \\
  & \omega & & & & -i \omega & & \\
  & & \begin{picture}(5,5) 
\multiput(0,4)(2,-2){3}{\makebox(0,0){.}}
\end{picture} & & & & \begin{picture}(5,5) 
\multiput(0,4)(2,-2){3}{\makebox(0,0){.}}
\end{picture} & \\
& & & \omega^{N-1} & & & & -i \omega^{N-1} \\
& & & \overline{\omega}^{N} & & & & 1 \\
  & & \begin{picture}(5,5) 
\multiput(0,4)(2,2){3}{\makebox(0,0){.}}
\end{picture} & & & & \begin{picture}(5,5) 
\multiput(0,4)(2,2){3}{\makebox(0,0){.}}
\end{picture} & \\
  & \overline{\omega}^2 & & & & i \overline{\omega}^2 & & \\
\overline{\omega} & & & & i \overline{\omega}& & & \\
\end{array}
\right)
\end{array}
$$
with $\omega=\exp(2 \pi i/4N)$. Equation~(\ref{dctIV}) is a consequence of Theorem~3.19 in
\cite{wickerhauser93} obtained by complex conjugation. 

\begin{theorem}
The discrete cosine transform $C_N^{\rm IV}$ and the discrete sine
transform $S_N^{\rm IV}$ can be realized with $O(\log^2 N)$ elementary
quantum gates; the quantum circuit for these transforms 
is shown in Figure~\ref{DCTIVcomplete}. 
\end{theorem}
\begin{figure*}[ht]
\input{dct4.pic}
\caption{\label{DCTIVcomplete}Complete quantum circuit for $\DCT_{\rm
IV}$}
\end{figure*}
\proof It remains to show that there exists an efficient quantum
circuit for the matrix $R_N$ in
equation~(\ref{dctIV}). 
A factorization of $R_N$ can be obtained as follows.  Denote by
$\overline{x}$ the one's complement of an $n$-bit integer $x$. 
We define a permutation matrix 
$\pi_1$ by $\pi_1\ket{0x}=\ket{0x}$
and $\pi_1\ket{1x}=\ket{1\overline{x}}$ for all integers $x$ in the
range of $0\le x< 2^n$. Denote by $D_1$ the diagonal matrix 
\[ 
D_1 = {\rm diag}(1, \omega, \dots, \omega^{N-1}, 
    \overline{\omega}^N, \dots, \overline{\omega}^2,
    \overline{\omega}).
\]
Then $R_N$ can be factored as 
$R_N = \pi_1\cdot D_1\cdot (\overline{B}\otimes \onemat_N).
$

Note that $\overline{B}\otimes \onemat_N$ is a single qubit operation,
and $\pi_1$ can be realized by controlled not operations. The
implementation of the diagonal matrix $D_1$ is more interesting. 
Note that
$$ 
\begin{array}{l@{\,\,}c@{\,\,}l}
\Delta_1=\diag(1,\omega,\dots,\omega^{N-1})&=&L_n\otimes\cdots\otimes L_2\otimes
L_1 \\
\Delta_2=\diag(\overline{\omega}^{N-1},\dots,\overline{\omega},1)&=&K_n\otimes\cdots
\otimes K_2\otimes K_1 
\end{array}
$$
where $L_j=\diag(1,\omega^{2^{j-1}})$ and
$K_j=\diag(\overline{\omega}^{\,2^{j-1}},1)$. 
Therefore, it is possible to write $D_1$ in the form 
$D_1=(C\otimes \onemat_N)\cdot (\Delta_1\oplus\Delta_2)$ with
$C=\diag(1,\overline{\omega})$. 
\nix{
The circuit for the diagonal matrix
$D_1$ is shown in Figure~\ref{diagonal}. 
\begin{figure*}[hb]
\input{dct4fact.pic}
\caption{\label{diagonal} Quantum circuit for the diagonal matrix $D_1$. }
\end{figure*}
}

The complete quantum
circuit for the DCT$_{\rm IV}$ is shown in
Figure~\ref{DCTIVcomplete}. Note that the last three single qubit
gates $C$, $B^\dagger$, and $M=\diag(e^{\pi i/4N},e^{\pi i/4N})$ can be
combined into a single gate $MB^\dagger C$.~\qed 

\section{DCT and DST of Type II}

The
implementation of the trigonometric transforms of type II follows a
similar pattern. Both transforms can be recovered from the DFT of
length 2N after multiplication with certain sparse matrices, 
cf.~Theorem~3.13 in~\cite{wickerhauser93}: 
\begin{equation}\label{dctII}
U_N^\dagger \cdot F_{2N} \cdot V_N = 
C_N^{\rm II} \oplus (-i) S_N^{\rm II},
\end{equation}
where
\[
V_N = 
\frac{1}{\sqrt{2}}
\left(
\begin{array}{rrrrrr}
1 & & & 1 & & \\
  & \begin{picture}(5,5) 
\multiput(0,4)(2,-2){3}{\makebox(0,0){.}}
\end{picture} & & & \begin{picture}(5,5) 
\multiput(0,4)(2,-2){3}{\makebox(0,0){.}}
\end{picture} & \\
 & & 1 &  & & \phantom{-} 1 \\
 & & 1 &  & & -1\\
  & \begin{picture}(5,5) 
\multiput(0,4)(2,2){3}{\makebox(0,0){.}}
\end{picture} & & & \begin{picture}(5,5) 
\multiput(0,4)(2,2){3}{\makebox(0,0){.}}
\end{picture} & \\
1 & & & -1 &  &
\end{array}
\right)
\]
and 

\[
\renewcommand{\arraystretch}{1.2}%
\setlength{\arraycolsep}{0.3em}%
U_N =
\left(
\begin{array}{cccccccc}
1 & & & & 0 & & & \\
  & \frac{\overline{\omega}}{\sqrt{2}} & & & 
  {-} \frac{i\overline{\omega}}{\sqrt{2}} & \begin{picture}(5,5)(-20,10)
\multiput(0,4)(2,-2){3}{\makebox(0,0){.}}
\end{picture} & & \\
  & & \begin{picture}(5,5) 
\multiput(0,4)(2,-2){3}{\makebox(0,0){.}}
\end{picture} & & & \begin{picture}(5,5) 
\multiput(0,4)(2,-2){3}{\makebox(0,0){.}}
\end{picture} & &\\
  & & & \frac{\overline{\omega}^{N-1}}{\sqrt{2}} & &
  & {-} \frac{i \overline{\omega}^{N-1}}{\sqrt{2}} & 0 \\
 & & &0 &  & & & -1\\
  & & \begin{picture}(5,5)(10,10) \multiput(0,0)(2,2){3}{\makebox(0,0){.}}
\end{picture} & \frac{\omega^{N-1}}{\sqrt{2}} & & & 
  \frac{i\omega^{N-1}}{\sqrt{2}} & \\
  & &  \begin{picture}(5,5) 
\multiput(0,0)(2,2){3}{\makebox(0,0){.}}
\end{picture} & & & \begin{picture}(5,5) 
\multiput(0,0)(2,2){3}{\makebox(0,0){.}}
\end{picture} & & \\
0 & \frac{\omega}{\sqrt{2}} & & & \frac{i\omega}{\sqrt{2}} & & &
\end{array}
\right),
\]
and $\omega=\exp(2\pi i/4N)$ with $i^2=-1$. 

\begin{figure*}[ht]
\input{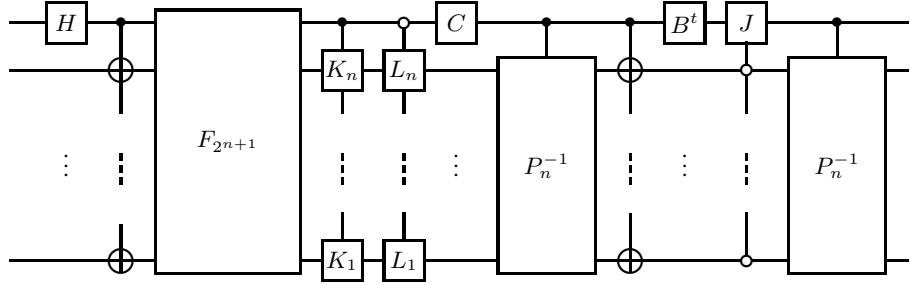}
\caption{\label{DCTIIcomplete}Complete quantum circuit for $\DCT_{\rm
II}$}
\end{figure*}

\begin{theorem}
The discrete cosine transform $C_N^{\rm II}$ and the discrete sine
transform $S_N^{\rm II}$ can be realized with $O(\log^2 N)$ elementary
quantum gates; the quantum circuit for these transforms 
is shown in Figure~\ref{DCTIIcomplete}. 
\end{theorem}
\proof We need to derive efficient quantum circuits for the matrices 
$V_N$ and $U_N$ in equation~(\ref{dctII}).
The matrix $V_N$ has a fairly simple decomposition in terms of quantum
circuits.

\begin{lemma} $V_N= \pi_1 ( H \otimes \onemat_N)$.
\end{lemma}
\proof It is clear that the Hadamard transform on the most
significant bit $H\otimes \onemat_N$ is -- up to a permutation of rows --
equivalent to $V_N$. The appropriate permutation of rows has been introduced in the previous
section, namely $\pi_1\ket{0x}=\ket{1x}$ and $\pi_1\ket{1x}=\ket{1\overline{x}}$ for
all $0\le x< 2^n$. We can conclude that $V_N=\pi_1( H \otimes
\onemat_N)$ as desired.~\qed
\medskip

The decomposition of $U_N$ is more elaborate. Notice that
$$
\begin{array}{l@{\;}l} 
U_N\ket{0\mathbf{0}}=\ket{0\mathbf{0}} \qquad U_N\ket{1\mathbf{1}}= (-1)\ket{1\mathbf{0}} \\[1ex]
U_N\ket{0 x}=\ds\frac{\overline{\omega}^{\,x}}{\sqrt{2}}\ket{0x}+
\frac{\omega^x}{\sqrt{2}}\ket{1x'}\\
U_N\ket{1y}=\ds -\frac{i\overline{\omega}^{\,y+1}}{\sqrt{2}}
\ket{0\,(y+1\bmod 2^n)}+\frac{i\omega^{y+1}}{\sqrt{2}}\ket{1\overline{y}} 
\end{array}
$$ 
for all integers $x$ in the range $1\le x< 2^n$ and all integers $y$
in $0\le y<2^n-1$. Here $\mathbf{0}$ and $\mathbf{1}$ denote the
$n$-bit integers $0$ and $2^n-1$ respectively. 

Define $D_0$ by $D_0\ket{1\mathbf{0}}=i\ket{1\mathbf{0}}$ and
$D_0\ket{x}=\ket{x}$ otherwise. We define a permutation $\pi_2$ by  
$\pi_2\ket{0x}=\ket{0x}$ and $\pi_2\ket{1x}=\ket{1(x+1\bmod 2^n)}$ for
all integers $x$ in $0\le x<2^n$. 
\begin{lemma}
$U_N=D_1^\dagger\, \overline{T}_N\, D_0^{\phantom{\dagger}}\, \pi_2$.
\end{lemma}
\proof
Since $D_1^\dagger\ket{0x}= \overline{\omega}^x\ket{0x}$
and  $D_1^\dagger\ket{1x}= \omega^{\,x'} \ket{1x}$, we obtain
$$ 
\begin{array}{l}
D_1^\dagger \overline{T}_N \ket{0x} = \ds \phantom{-}
\frac{\overline{\omega}^x}{\sqrt{2}}\ket{0x} + 
\frac{\omega^x}{\sqrt{2}}\ket{1x'}\\[2ex]
D_1^\dagger \overline{T}_N \ket{1x} = \ds 
-\frac{i\overline{\omega}^x}{\sqrt{2}}\ket{0x} + 
\frac{i\omega^x}{\sqrt{2}}\ket{1x'}
\end{array}
$$
We have $D_0\pi_2\ket{0x}= \ket{0x}$ and moreover $D_0\pi_2\ket{1x}=\ket{1
(x+1\bmod 2^n)}$ for all integers $x$ in $0\le x<2^n-1$, and
$D_0\pi_2\ket{1\mathbf{1}}=i\ket{1\mathbf{0}}$. We note that
$(x+1\bmod 2^n)'=\overline{x}$, whence
combining $D_1\overline{T}_N$ with
$D_0\pi_2$ shows the result.~\qed

Recall that $T_N=\pi D$. It follows that 
$$ U_N^\dagger = \pi_2^{-1} (\overline{D}_0 D^t) \pi^{-1}D_1.$$
The implementation of $D_1$ has been described in the section on the
DCT$_{\rm IV}$, and the implementation of $\pi$ (and hence $\pi^{-1}$)
is contained in the section on the DCT$_{\rm I}$. The implementation
of $\pi_2^{-1}$ is also straightforward. It remains to find an
implementation of $\overline{D}_0D^t$. We observe that
$$
\begin{array}{l@{\,}c@{\,}l@{\;}l@{\,}c@{\,}l}
\overline{D}_0D^t\ket{0\mathbf{0}}&=&\ket{0\mathbf{0}}, &
\overline{D}_0D^t\ket{0x}&=&\frac{1}{\sqrt{2}}\ket{0x}{+}\frac{i}{\sqrt{2}}\ket{1x},\\[1
ex]
\overline{D}_0D^t\ket{1\mathbf{0}}&=&{-}i\ket{1\mathbf{0}}, &
\overline{D}_0D^t\ket{1x}&=&\frac{1}{\sqrt{2}}\ket{0x}{-}\frac{i}{\sqrt{2}}\ket{1x}.
\end{array}
$$\renewcommand{\arraystretch}{1.0}%
This can be accomplished by a single qubit operation followed by a 
multiply conditioned gate, where the single qubit operation is given by 
$B^t\otimes
\onemat_N$ 
and the conditional gate acts via 
$$J =\ds\frac{1}{\sqrt{2}}\mat{1}{-i}{-i}{1}.$$ 
The full circuit is shown in Figure~\ref{DCTIIcomplete}.
The statement about the complexity is clear.~\qed

\section{Conclusions}
Signal processing methods have proved to be useful in virtually all
known quantum algorithms. A basic problem in the design of quantum
algorithms is to choose a well-adapted basis to recoup relevant
information about the state of the system. The well-known
decorrelation properties of DCTs may prove to be useful within this
framework. We have shown that the DCT and DST of types I, II, III, and
IV can be realized with a polylogarithmic number of elementary
operations on a quantum computer. Compared to the classical
realization of the DCT, this is a tremendous speed-up, making the DCT
attractive in the design of other quantum algorithms. 

\medskip\noindent \textbf{Acknowledgments.} We thank
Hyunyoung Lee for comments that improved the 
presentation of this paper. We thank the European 
Community for supporting this research under IST-1999-10596 (Q-ACTA).


\begin{thebibliography}{10}

\bibitem{Shor:94}
P.~W. Shor,
\newblock ``{Algorithms for Quantum Computation: Discrete Logarithm and
  Factoring},''
\newblock in {\em Proc. FOCS 94}. 1994, pp. 124--134, IEEE Computer Society
  Press.

\bibitem{cirac95}
J.I. Cirac and P.~Zoller,
\newblock ``Quantum computations with cold trapped ions,''
\newblock {\em Phys. Rev. Lett.}, vol. 74, no. 20, pp. 4091--4094, 1995.

\bibitem{nagerl98}
H.~N{\"a}gerl, D.~Leibfried, H.~Rohde, J.~Thalhammer, G.~Eschner,
  F.~Schmidt-Kaler, and R.~Blatt,
\newblock ``Laser addressing of individual ions in a linear ion trap,''
\newblock {\em Physical Review A}, vol. 60, no. 1, pp. 145--148, 1999.

\bibitem{kane98}
B.E. Kane,
\newblock ``A silicon-based nuclear spin quantum computer,''
\newblock {\em Nature}, vol. 393, no. 6681, pp. 133--137, 1998.

\bibitem{imamoglu99}
A.~Imamo{\-{g}}lu, D.D. Awschalom, G.~Burkard, D.P. DiVincenzo, D.~Loss,
  M.~Sherwin, and A.~Small,
\newblock ``Quantum information processing using quantum dot spins and
  cavity-qed,''
\newblock {\em Phys. Rev. Lett.}, vol. 83, pp. 4204--4207, 1999.

\bibitem{rauschenbeutel99}
A.~Rauschenbeutel, G.~Nogues, S.~Osnaghi, P.~Bertet, M.~Brune, J.M. Raimond,
  and Haroche S.,
\newblock ``Coherent operation of a tunable quantum phase gate in cavity
  {QED},''
\newblock {\em Phys. Rev. Lett.}, vol. 83, no. 24, pp. 5166--5169, 1999.

\bibitem{knill01}
E.~Knill, R.~Laflamme, and G.J. Milburn,
\newblock ``A scheme for efficient quantum computation with linear optics,''
\newblock {\em Nature}, vol. 409, pp. 46--52, 2001.

\bibitem{nist01}
D.~Kielpinski~et al.,
\newblock ``Recent results in trapped-ion quantum computing at {NIST},''
\newblock {\em To appear in Proc. of IQC 2001}, 2001.

\bibitem{RY:90}
K.~R. Rao and P.~Yip,
\newblock {\em Discrete Cosine Transform: Algorithms, Advantages, and
  Applications},
\newblock Academic Press, 1990.

\bibitem{BBCshort:95}
A.~Barenco~{\em{}et~al.},
\newblock ``{Elementary gates for quantum computation},''
\newblock {\em Physical Review~A}, vol. 52, no. 5, pp. 3457--3467, Nov. 1995.

\bibitem{wickerhauser93}
V.~Wickerhauser,
\newblock {\em Adapted Wavelet Analysis from Theory to Software},
\newblock A.K. Peters, Wellesley, 1993.

\bibitem{PRBshort:99}
M.~P{\"u}schel, M.~R{\"o}tteler, and Th. Beth,
\newblock ``{Fast Quantum Fourier Transforms for a Class of non-abelian
  Groups},''
\newblock in {\em Proc. AAECC-13}. 1999, vol. 1719 of {\em LNCS}, pp. 148--159,
  Springer.

\end{thebibliography}
\end{document}